\documentclass[twocolumn
]{aastex63}

\usepackage{amssymb}
\usepackage{amsmath}
\usepackage{txfonts}

\usepackage{natbib}

\usepackage{hyperref}
\hypersetup{linkcolor=red,citecolor=blue,urlcolor=cyan,colorlinks=true}

\usepackage{graphicx}
\usepackage{dcolumn}
\usepackage{bm}
\usepackage{color}
\usepackage{soul}
\usepackage{mathtools}
\usepackage[T1]{fontenc}

\def\aap{A\&A}

\def\apjl{ApJL}
\def\apjs{ApJS}

\date{\today}

\shorttitle{Crustal failure as a tool to probe hybrid stars}
\shortauthors{Pereira, Bejger, Haensel and Zdunik}
\graphicspath{{./}{figures/}}

\begin{document}

\title{Crustal failure as a tool to probe hybrid stars}

\correspondingauthor{Jonas P.~Pereira}
\email{jpereira@camk.edu.pl}

\author{Jonas P.~Pereira}
\affiliation{Nicolaus Copernicus Astronomical Center, Polish Academy of Sciences, Bartycka 18, 00-716, Warsaw, Poland}

\author{Micha{\l} Bejger}
\affiliation{INFN Sezione di Ferrara, Via Saragat 1, 44122 Ferrara, Italy}
\affiliation{Nicolaus Copernicus Astronomical Center, Polish Academy of Sciences, Bartycka 18, 00-716, Warsaw, Poland}

\author{Pawe{\l} Haensel}
\affiliation{Nicolaus Copernicus Astronomical Center, Polish Academy of Sciences, Bartycka 18, 00-716, Warsaw, Poland}

\author{Julian Leszek Zdunik}
\affiliation{Nicolaus Copernicus Astronomical Center, Polish Academy of Sciences, Bartycka 18, 00-716, Warsaw, Poland}
 
\begin{abstract}
It is currently unknown if neutron stars (NSs) are composed of nucleons only or are {\it hybrid stars}, i.e., in addition to nucleonic crusts and outer cores, they also possess quark cores. Quantum chromodynamics allows for such a possibility, but accurate calculations relevant for compact stars are still elusive. Here we investigate some crust-breaking aspects of hybrid stars. We show that the crust-breaking  
frequency and maximum fiducial ellipticity are sensitive to the quark-hadron density jump and equation of state stiffness. Remarkably, the crust-breaking frequency related to static tides scales linearly with the mass of the star (for a given companion's mass), and its slope encompasses information about the microphysics of the star. However, for precise crust-breaking frequency predictions, relativistic corrections to Kepler’s third law and the Newtonian tidal field should not be ignored. When a liquid quark core touches an elastic hadronic phase (the result of a significant energy-density jump), the maximum ellipticity can increase around an order of magnitude when compared to a liquid quark core touching a liquid hadronic phase.
That is relevant because it would increase the odds of detecting continuous gravitational waves from NSs. Our order-of-magnitude analysis also suggests that a given upper limit to the ellipticity(crust-breaking frequency) could have representatives in stars with either small or intermediate(large) energy-density jumps.
Therefore, when upper limits to the ellipticity for isolated stars are better constrained or electromagnetic radiation (e.g., gamma-ray precursors) is detected along with gravitational waves in inspiraling binary systems, they may help constrain some aspects of phase transitions in NSs.

\end{abstract}


\section{Introduction}

The direct detection of gravitational waves (GWs) from binary systems of black holes (BHs) \citep{PhysRevLett.116.061102} and neutron stars (NSs) \citep{2017PhRvL.119p1101A} allowed to test the properties of these astrophysical sources. As more events are detected, statistical studies also become possible, which could reduce uncertainties for GW observables. While GW observations from binary BH systems are more common in the LIGO-Virgo transient GW catalogs \citep{2019PhRvX...9c1040A,2020arXiv201014527A,2021arXiv210801045T, 2021arXiv211103606T} due to their intrinsically larger GW strain amplitude, NS measurements are also present. The confident detection of GWs from the GW170817 event \citep{2017PhRvL.119p1101A}, with a multi-messenger counterpart of GRB170817A and a subsequent kilonova \citep{2017ApJ...848L..12A} has furnished us with first constraints on the tidal deformations of NSs, which in turn allowed for significant restrictions on several macroscopic and microscopic aspects of the dense matter equation of state (EOS). 

The detection of other types of sources, related to long-duration GWs (continuous waves, CWs) remains a possibility; for recent LIGO-Virgo-KAGRA searches, see, e.g., \citet{2019PhRvD.100b4004A,2021PhRvD.103f4017A,2022arXiv220100697T,2022ApJ...935....1A}. CWs naturally contrast with transient GWs, such as the GW170817 event, and are much more challenging to be detected due to their smaller GW strain amplitude. The non-detection allows however the setting of upper limits to the GW strain, which has already surpassed the physically interesting limits, like the spin-down limit for non-axisymmetric rotating NSs for some targets \citep{2022ApJ...935....1A}. Although this may not be restrictive enough, it is noteworthy from the observational and data analysis viewpoints. Current upper limits to the GW strain can be immediately translated into upper limits to the \textit{fiducial ellipticity} 
of rotating NSs  (whose reference value for their principal moment of inertia $I_{zz}$ is $10^{45}$ g cm$^2$) when the distance to the source is known \citep{2022arXiv220100697T,2022ApJ...935....1A}, a measure of how much the objects are deformed in a quadrupolar fashion, to the leading order approximation. This quantity already gives us complementary information on their interiors [and also on the shape of a star's surface \citep{2013PhRvD..88d4016J}], different from those obtained by binary inspirals.

Associated with the maximum ellipticity is the \textit{breaking strain} of an elastic/solid phase of an NS, beyond which it starts failing as it cannot sustain a larger deformation. The breaking strain  is directly related to the maximum value of the scalar strain \citep{2019CQGra..36j5004A}, an invariant quantity that in general depends on the microphysical aspects of a star (and also on the star's evolutionary history). In this sense hybrid NSs, i.e., stars with a quark (exotic) phase and a hadronic phase, is an interesting possibility to investigate. In the following, for simplicity, we will denote compact stars with hybrid interiors as NSs, adding clarifications if necessary. One would expect the scalar strain of a hybrid star to depend on its phase transition aspects for several reasons. Mathematically, that could be the case because further boundary conditions would be inserted into the problem, influencing strongly deformations and their outcomes such as ellipticities \citep{2000MNRAS.319..902U,2006MNRAS.373.1423H,2013PhRvD..88d4004J,2020arXiv200310781P,2020PhRvD.101j3025G,2021MNRAS.500.5570G,2022arXiv220903222M}. Physically, phase transitions could shorten the hadronic density range in NSs, rendering them easier to deform; they can also lead to harder-to-break NSs due to the increase of their compactness associated with the softening of the dense-matter EOS. 
Finally, even if the same breaking strain condition is taken for a given region of hybrid and non-hybrid stars (e.g., associated with their crusts/hadronic phases), the different behaviors of the scalar strain in each case would lead to different fracturing consequences. To the best of our knowledge, these issues have not been explored yet in great detail.  

In the context of inspiraling binary systems, one could also estimate where in the orbit the elastic crust of an NS would start breaking due to tidal interactions. Studies for hadronic NSs with selected EOSs suggest that they may not break before merger \citep{2020PhRvD.101j3025G}, but the literature lacks similar studies for hybrid stars and for a larger variety of EOSs. We partially fill this gap here by means of order-of-magnitude estimates and general trends hybrid stars could present. The motivation is the possibility of other types of observables, such as electromagnetic (EM) radiation before the merger, related to precursors of short gamma-ray bursts (see, e.g., \citet{2020ApJ...902L..42W,2020PhRvD.102j3014C,2021Galax...9..104W} and references therein), which could be explained in a variety of ways associated with crustal failure \citep{2012PhRvL.108a1102T,2020PhRvD.101h3002S,2021MNRAS.508.1732K,2021MNRAS.504.1273P,2022MNRAS.513.4045K,2022MNRAS.514.1628K}.

We organize this paper in the following way. In Sec. \ref{sec:yielding_strain}, we describe the formalism for the crustal failure of an NS. Sections \ref{sec:ellipticity} and \ref{sec:breaking_frequencies} explain aspects concerning maximum ellipticities of (isolated) NSs and the crust-breaking frequencies of NSs in binaries. The NS models investigated are explained in Sec. \ref{sec:models}. In Sec. \ref{sec:results} we present the main results for the crust-breaking frequency and maximum ellipticity of hybrid NSs and discuss them in Sec. \ref{sec:conclusions}. Unless otherwise stated, we work with geometric units.

\section{Breaking strain}
\label{sec:yielding_strain}
The breaking strain of a solid is a nontrivial issue, involving complex physics beyond the elastic regime that can only be probed with many body simulations. The state-of-art simulations of \citet{PhysRevLett.102.191102} point to a break of the lattice structure when the scalar strain $\Theta$ is larger than $0.1$. It is defined (using the von Mises criterion) as  \citep{2019CQGra..36j5004A}
\begin{equation}
    \Theta \equiv \frac{1}{2\check{\mu}}\sqrt{\frac{3}{2}\Delta\pi^{ab}\Delta\pi_{ab}}\label{scalar_strain},
\end{equation}
where $\Delta \pi^{b}_a$ is the Lagrangian perturbation of the anisotropic stress tensor, given by \citep{2011PhRvD..84j3006P,2019CQGra..36j5004A,2021MNRAS.507..116G}
\begin{equation}
\Delta\pi^b_a = -\check{\mu} \left({\cal P}^c_a{\cal P}^{db}-\frac{1}{3}{\cal P}_a^b{\cal P}^{cd}\right)\Delta g_{cd} \label{delta_pi},
\end{equation}
with $\check{\mu}$ the shear modulus of the lattice, ${\cal P}_{ab}\equiv g_{ab}+u_au_b$ denoting the projection onto the orthogonal directions of the fluid's four-velocity $u^a$, $g_{ab}$ the background spacetime, and $\Delta g_{cd}$ the Lagrangian perturbation of the metric (for further details and assumptions, see \citet{2020arXiv200310781P,2021MNRAS.507..116G}). 

A scalar strain larger than $0.1$ would mark the beginning of the failing process of a lattice, but this value depends on many assumptions. Other studies suggest that an elastic lattice would start failing at even smaller $\Theta$, for instance, $\Theta=0.01$ \citep{2010MNRAS.407L..54C,2018MNRAS.480.5511B}, in addition to the critical strain being anisotropic in general \citep{2018MNRAS.480.5511B}. Thus, in order to have upper limits to physical quantities and to focus on the simplest case, we take that elastic crusts/elastic hadronic phases start breaking when $\Theta > 0.1$.

In this work we want to find the influence a sharp phase transition \citep{2018ApJ...860...12P} could have on the breaking of the crust/hadronic phase of a hybrid star (isolated or in a binary system). We follow the approach of \citet{2020PhRvD.101j3025G,2020arXiv200310781P} to find the solutions for static perturbations in general relativity. They are necessary for calculating $\Theta$ in the case of maximum ellipticities and crust-breaking frequencies. More specifically, one needs to solve the equations associated with spacetime and fluid perturbations in the presence of elasticity, while respecting  the appropriate boundary conditions in this case \citep{2020arXiv200310781P}. The background spacetime is obtained from the solution of the Tolman-Oppenheimer-Volkoff (TOV) equations with a given EOS for the NS matter. For all the equations and their solution strategy, see \citet{2020PhRvD.101j3025G,2020arXiv200310781P}.

\section{Ellipticities}
\label{sec:ellipticity}
In Newtonian dynamics for spherically symmetric backgrounds, the perturbation's multipole moment is defined as \citep{2000MNRAS.319..902U}
\begin{equation}
    Q_{lm}\equiv \int \delta \rho_{lm}(r)r^{l+2}dr\label{quadrupole_moment},
\end{equation}
where the mass density $\delta\rho_{lm}$ is such that $\delta \rho (r,\theta, \phi) \equiv \sum_{l,m}\delta\rho_{lm}(r)Y_{lm}(\theta,\phi)$, with  $Y_{lm}(\theta,\phi)$ the spherical harmonics. For spherically symmetric backgrounds, $\delta \rho_{lm}\equiv \delta \rho_l$--the radial part of the modes degenerates for the different $m$ related to a given $l$, so all their $Q_{lm}$ are the same. Due to its relevance for GW emission, from now on we focus on $l=m=2$.

In the case of general relativity, Eq. \eqref{quadrupole_moment} becomes ambiguous due to the coordinate system freedom. As in the case of tidal deformations, the consistent way of finding $Q_{22}$ is to use the asymptotic expansion of the $g_{tt}$ metric component and its relation to the multipole moments and the tidal field. As shown in \citet{2013PhRvD..88d4004J,2008ApJ...677.1216H,2021MNRAS.507..116G} in the axisymmetric case (the normalization is chosen in a way to coincide with the definitions of \citet{2000MNRAS.319..902U}),
\begin{equation}
    Q_{22}=\frac{M^3c_1}{\pi}\label{quadrupole_relativity}
\end{equation}
with  $M$ the star's mass and $c_1$ an arbitrary constant associated with the external solution to $H_0$ [related to the $tt$-metric perturbation in the Regge-Wheeler gauge] \citep{2008ApJ...677.1216H}:
\begin{equation}
    H_0= c_1Q_2^2\left( \frac{r}{M}-1\right) + c_2P_2^2\left( \frac{r}{M}-1\right)  \label{H0_external},
\end{equation}
where $Q_2^2$ and $P^2_2$ are the associate Legendre polynomials. The asymptotic form of $H_0$ shows that $c_1$ is related to the body's response to a tidal field \citep{2008ApJ...677.1216H}, as Eq. \eqref{quadrupole_relativity} clearly suggests. Thus, to obtain $Q_{22}$, one needs to find $H_0$ inside the star and use boundary conditions on its surface to get $c_1$. It carries, in a global way, all the internal information of the star. That also includes the effect of boundary conditions that depend on possible phase transitions. In particular, energy density jumps (first-order phase transitions) could leave an imprint on $c_1$. 

The ellipticity itself is defined as \citep{2005PhRvL..95u1101O,2013PhRvD..88d4004J}
\begin{equation}\label{ellipticity_star}
    \varepsilon \equiv \sqrt{\frac{8\pi}{15}}\frac{Q_{22}}{I_{zz}},
\end{equation}
where $I_{zz}$ is the principal moment of inertia of the NS, whose fiducial value is $10^{45}$~g cm$^2$. As the physically important quantity for GW emission is the quadrupole moment of deformed stars, we use the fiducial moment of inertia in our ellipticity calculations. With the fiducial ellipticity, one can readily calculate $Q_{22}$.

To compute $\varepsilon$, one also needs a relaxed reference configuration. The spherically symmetric, perfect-fluid configurations may not be the most appropriate one \citep{2021MNRAS.500.5570G,2021MNRAS.507..116G}. A natural configuration could be axisymmetric due to its expected high rotation after, e.g., the birth, or merger of two NSs. A deformed, non-rotating perfect-fluid configuration due to given forces also seems a reasonable choice for the relaxed reference state. By subtracting off this contribution ($Q_{22}^{\rm{ref}}$) from the total $Q_{22}^{\rm{total}}$, one would end up with a ``net'', almost force-independent quadrupole moment \citep{2021MNRAS.500.5570G,2021MNRAS.507..116G,2022arXiv220903222M}, to be linked with the degree of deformation sustained by the star. Thus, we assume that $Q_{22}\equiv |Q_{22}^{\rm{total}}-Q_{22}^{\rm{ref}}|$, and that the NS ellipticity is given by Eq. \eqref{ellipticity_star}. The expression for $\Theta$ for $l=m=2$, from which the maximum ellipticity can be calculated, can be found in \citet{2021MNRAS.507..116G}.

\section{Crust-breaking frequencies}
\label{sec:breaking_frequencies}
We assume a hybrid NS in a binary system. Its companion could be another star or a black hole. The orbital frequency at which the crust starts yielding is related to a critical distance to its companion since the strain, present due to tidal forces,  increases with decreasing distance to the companion. Kepler's third law connects the distance between the centers of mass of the two objects in a binary system and the orbital frequency (in Sec. \ref{sec:results} we estimate the errors associated with this assumption). 

A relevant point for a spherically-symmetric NS is which $m$ to take to calculate the scalar strain and hence find the consequences of its maximum value. For a given $l$, one would expect that all $m$ components are excited. Therefore, it seems reasonable to assume that the most relevant $m$ for the fracture of a crust is the one that leads to the smallest orbital frequency (largest separation) for a given $\Theta$. It turns out that it is $m=0$. For the expression of $\Theta$ in this case, see \citet{2020PhRvD.101j3025G}. 
We quote where in the binary's evolution the crust breaking takes place by using the dominant mode of GWs ($l=m=2$), whose frequency is twice the orbital frequency.

\section{Models}
\label{sec:models}
To gain intuition on the general properties of orbital frequencies and ellipticities associated with a crust breaking, we start our analysis with toy-model EOSs. They are rough representations of properties expected for a quark and a hadronic phase in a hybrid star. For the quark phase, we assume a simple linear relationship  (MIT-bag-like model) between the pressure $p$ and the energy density $\rho$ with free parameters being the speed of sound $c_s$ and a critical density $\rho^*$ where the pressure is null. In the MIT bag model this constant is related to the bag constant $B$ by $\rho^*=4B$, but in our case it is just a parameter of a linear EOS for the quark phase. This EOS is of the form
\begin{equation}
    p(\rho)=c_s^2(\rho-\rho^*)\label{EOS_quark}.
\end{equation}
We choose to fix $\rho^*$ by already known constraints on the pressure at twice the saturation density $p(2\rho_{\rm{sat}})$ coming from GW observations \citep{2018PhRvL.121p1101A}. In other words, from Eq. \eqref{EOS_quark}, $\rho^*=2\rho_{\rm{sat}}-p(2\rho_{\rm{sat}})/c_s^2$. Therefore, for the quark phase, the only free parameter is the speed of sound. The conclusions based on multimessenger astronomy with Bayesian analyses suggest a range of sound speeds for the quark phase in hybrid stars, even including stiff models \citep{2021PhRvC.103c5802X,2021ApJ...913...27L,2021arXiv210605313L}. They favor  $c_s^2\simeq 1$, indicating that the quark EOS should be stiff so that hybrid stars with masses around 1.4 $M_{\odot}$ would exist. 

\begin{figure} 
   \includegraphics[width=\columnwidth]{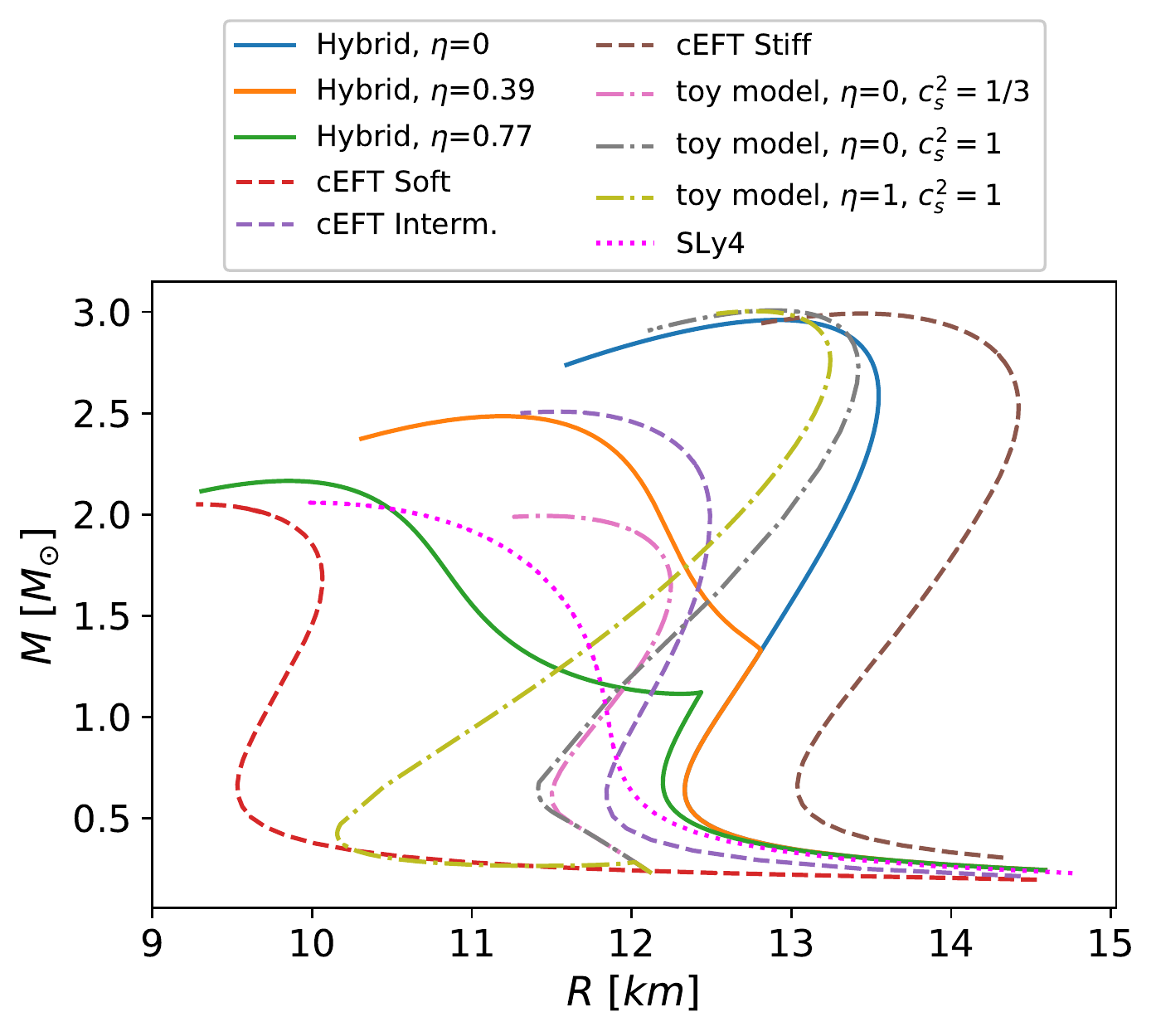}\caption{Mass-radius relations for the EOSs used in our analysis. We take the soft, intermediate, and stiff cEFT EOSs as representative EOSs for one-phase stars. For the hybrid NS models, we take those of \citet{2020arXiv201106361P} with density jumps $\eta=(0, 0.39, 0.77)$, covering strong and weak phase transitions (for details, see, e.g., \citet{2019A&A...622A.174S} and references therein). For these models, the SLy4 EOS is joined with a polytropic EOS (whose adiabatic index is 4.5) at the densities ($0.21,0.21,0.235$) fm$^{-3}$, respectively. The polytropic EOS extends up to 0.335 fm$^{-3}$ $(\sim 2.1\rho_{\rm{sat}})$, marking the end of the hadronic phase. For the quark phase, a simple linear (MIT bag-like) EOS with $c_s^2=1$ is taken. For further details, see Fig. 1 of \citet{2020arXiv201106361P}. Toy-model EOSs are also shown for completeness for some speeds of sound ($c_s^2$) and $\eta$ values.}\label{fig:MRtotal}
   \end{figure}

For the hadronic phase of our toy-model EOS, we choose a polytropic model 
\begin{equation}
    p=K\rho^2, \label{EOS_hadron}
\end{equation}
with $K=100$ km$^2$, leading to reasonable values of radius at $1.4$ $M_{\odot}$ and the maximum mass. As can be seen in Fig. \ref{fig:MRtotal} for some examples of these toy-model EOSs, the associated hybrid stars fulfill some basic criteria coming from EM constraints, which roughly state that 1.4 $M_{\odot}$ and 2.0 $M_{\odot}$ stars should have radii around 12 km \citep{2019ApJ...887L..24M,2019ApJ...887L..21R,2021ApJ...918L..28M,2021arXiv210506980R}. In addition, the  (dimensionless) tidal deformation/deformability [$\Lambda\equiv 2/3(M/R)^{-5}k_2$, where $M$ is the star's mass, $R$ is the star's radius and $k_2$ is the Love number \citep{2008ApJ...677.1216H,2009PhRvD..80h4035D,2009PhRvD..80h4018B}] of 1.4 $M_{\odot}$ hybrid stars with the above EOSs are in the range 450-550 for $1/3\leq c_s^2\leq 1$, in agreement with GW constraints \citep{2019PhRvX...9a1001A}. 

To add a further degree of realism to our estimates, we also consider other EOSs \citep{2020arXiv200310781P, 2020arXiv201106361P}. We assume the SLy4 EOS \citep{2001A&A...380..151D} at crust densities, a ``hadronic'' core phase approximated by polytropic EOSs, and a simple MIT bag-like model for the inner quark core, with the speed of sound equal to unity. We also employ the chiral effective field theory (cEFT) EOSs with varying stiffness \citep{Hebeler:2013nza,2020ApJ...901..155G} as benchmark EOSs without phase transitions. Since we are interested in the elastic properties of the crust, which comprises the low-density region of NSs, these EOSs are a reasonable set for our analyses. Their mass-radius relations are shown in Fig. \ref{fig:MRtotal}.

We assume that hybrid stars have a sharp exotic-hadron phase transition (Maxwell construction), thus characterized by a first-order phase transition, which presents an energy density jump $\eta \equiv \rho_{q}/\rho_{h}-1$, with $\rho_q$ the energy density at the top of the exotic phase (e.g., quark) and $\rho_h$ the energy density at the bottom of the hadronic phase. Strictly speaking, we do not work with self-bound quark stars \citep{1986A&A...160..121H,1986ApJ...310..261A,2010PhRvD..82b4016P}. Although interesting, we ignore in this work other constructions, such as the ones leading to mixed states \citep{2022PhRvD.105l3015P}. It turns out that the mixed state is indistinguishable from the Maxwell construction when the surface tension is larger than a critical quantity \citep{2019PhRvC.100b5802M}, and both are currently unknown and EOS dependent. Thus, in this first approach, it is natural to choose the simplest construction, which is Maxwell's. We take elasticity at the level of perturbations and assume that it is constrained to only the star's hadronic part (crust). We choose $\rho=1.5\times 10^{14}$ g cm$^{-3}$ as the density at the base of the elastic crust \citep{2011PhRvC..83d5810D}. (Thus, it is the maximum density in the elastic part of the crust.) For densities larger than $\rho=1.5\times 10^{14}$ g cm$^{-3}$, when present, the crust/hadronic phase is liquid. The density at the top of the elastic crust is taken to be $10^{7}$ g cm$^{-3}$, where the NS ocean starts \citep{2020PhRvD.101j3025G,2020arXiv200310781P}.

A shear modulus of the hadronic crust is necessary to solve the perturbation equations with elasticity. We work with phenomenological models that capture the main aspects of first-principle calculations and those coming directly from EOSs (see, e.g., \citet{2008A&A...491..489Z}). Since we are interested in general trends for more precise future analysis, for the phenomenological shear moduli we take $\check{\mu}= \kappa_{\rho}\rho$ \citep{2006MNRAS.373.1423H,2008LRR....11...10C}, where $\rho$ is the energy density. For numerical purposes, we use $\kappa_{\rho}=10^{16}$ cm$^{2}$s$^{-2}$ since it would be a fit compatible with $\check{\mu}= \kappa_{p} p$ with $\kappa_p\approx 0.01$ \citep{2008LRR....11...10C}, for the EOSs used. For completeness, we stress that elasticity could also be present in other parts of an NS, such as its mixed phase/state \citep{2012PhRvD..86f3006J,2013PhRvD..88d4004J}, or even the quark core \citep{2007PhRvD..76g4026M}. However, we leave breaking analyses of these phases for future works.

The adiabatic index of dense matter is also relevant for getting the critical breaking frequencies. If taken as the equilibrium one, the star would become more deformable and hence easier to break around the neutron drip density $\rho_{\rm{drip}}$  \citep{2020PhRvD.101j3025G}. However, when perturbations are present, the adiabatic index could be very different from the equilibrium one, mainly near $\rho_{\rm{drip}}$  \citep{1977ApJ...217..799C}. That can play a role in tidal deformation calculations, especially for the case of elasticity, meaning that one should ignore fracturing effects around $\rho_{\rm{drip}}$ if one uses an equilibrium adiabatic index. For catalyzed matter, an equilibrium adiabatic index means that all reactions between nuclei leading to the minimum of energy are allowed, which can be true only for high temperatures and/or densities. To obtain the breaking frequency associated with most of the crust for realistic EOSs, we search for the points that break at densities much larger than $\rho_{\rm{drip}}$ ($\rho>2\times 10^{13}$ g cm$^{-3}$, just to be roughly two orders of magnitude larger than the neutron-drip density). This means that to a good approximation ($\lesssim 10\%$ \citep{1977ApJ...217..799C}), we can take the adiabatic index as the equilibrium one \citep{1977ApJ...217..799C} and simply ignore the its particularities around the neutron drip density and smaller densities, or even in other physical situations (see, e.g., \citet{2020MNRAS.491.1064G} for some consequences of a frozen adiabatic index and \citet{2022A&A...665A..74F} when accretion is involved). We plan to carry out more precise analysis on the crust-breaking frequency with the exact frozen adiabatic index in future work.

\section{Results}
\label{sec:results}
We start our analysis with the toy-model EOSs. Quark-hadron $\eta$s 
are roughly $0\lesssim\eta\lesssim 2$ from Bayesian analysis with current constraints (see, e.g., \citet{2021ApJ...913...27L,2021PhRvD.103f3026T,2021PhRvC.103c5802X} and references therein). However, larger values for $\eta$ may also be possible. To cover all possibilities, we take $\eta$ as a free parameter and allow it to be as large as $10^3-10^4$. When working with such models, our primary motivation is to gain intuition on possible outcomes for more realistic analyses. Regarding GW frequencies, \citet{2020PhRvD.101j3025G} have estimated (within the Newtonian context) the maximum values before the merger to be around 2kHz for stars with $M=1.4 M_{\odot}$ and $R=10$ km. We take this as a rough reference frequency for the yield of the crust. In order to have some contact with realistic EOSs concerning crust-breaking aspects, for toy-model EOSs we search for the breaking strain for densities larger than $\rho_{\rm drip}$.

\begin{figure} 
   \includegraphics[width=\columnwidth]{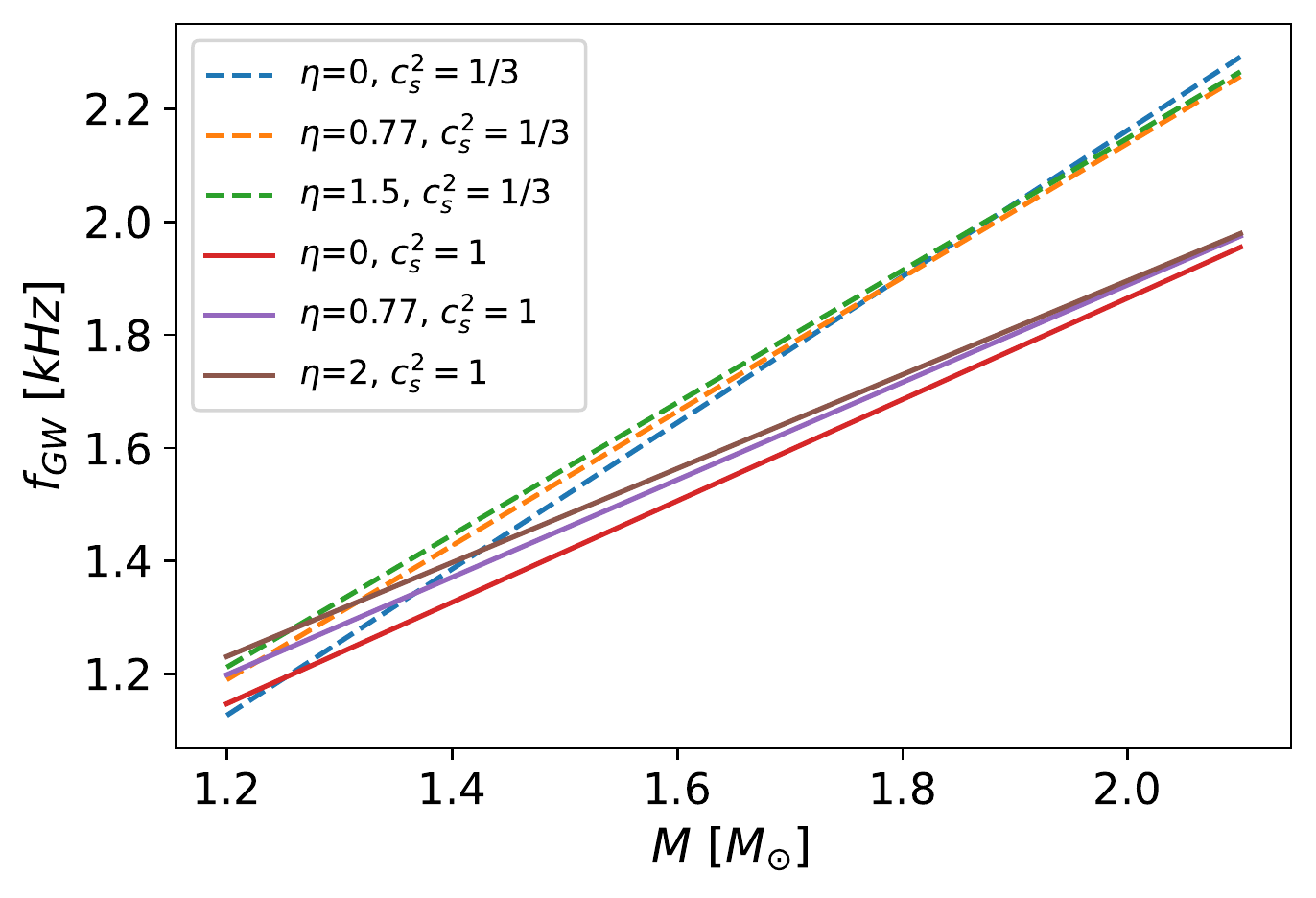}   
   \caption{Crust-breaking GW frequency of NSs in a binary system with a $1.4M_{\odot}$ companion for different density jumps and speeds of sound in the quark phase. For the shear modulus, we take the simple model $\check{\mu}=\kappa_{\rho}\rho$, with $\kappa_{\rho}=10^{16}$ cm$^{2}$s$^{-2}$. Straight lines are very good fits for the crust-breaking frequencies based on the numerical integrations for quasi-static perturbations.}
   \label{fig:MfBookKeeping}
   \end{figure}

Figure \ref{fig:MfBookKeeping} shows the behavior of the critical GW breaking frequency when the crust starts to fail as a function of the mass of the hybrid star for toy-model EOSs when the binary companion has a 1.4 $M_{\odot}$. The relationship is linear (deviations are up to $2\%-3\%$ for $c_s^2=1/3$ and $0.2\%-0.3\%$ for $c_s^2=1$) for masses not too close to the maximum mass. It depends on the stiffness of the EOS and, more weakly, on the density jump $\eta$. The linear relationship is remarkable because the breaking depends on many nontrivial aspects of boundary conditions for perturbations. The increase (decrease) of the crust-breaking frequency with the mass (speed of sound) is expected, because the star becomes more (less) compact and hence more (less) difficult to be deformed. For masses around 1.2 $M_{\odot}$, the critical frequencies are similar for small and large $c_s^2$, whereas the difference becomes more distinct for larger masses. In addition, the critical frequencies get less dependent on $\eta$ for larger masses. The above suggests that for some range of masses, one could obtain information about the stiffness of the quark EOS using events associated with the NS crust breaking in a binary system. We return to this issue in the discussion section.

\begin{figure} 
   \includegraphics[width=\columnwidth]{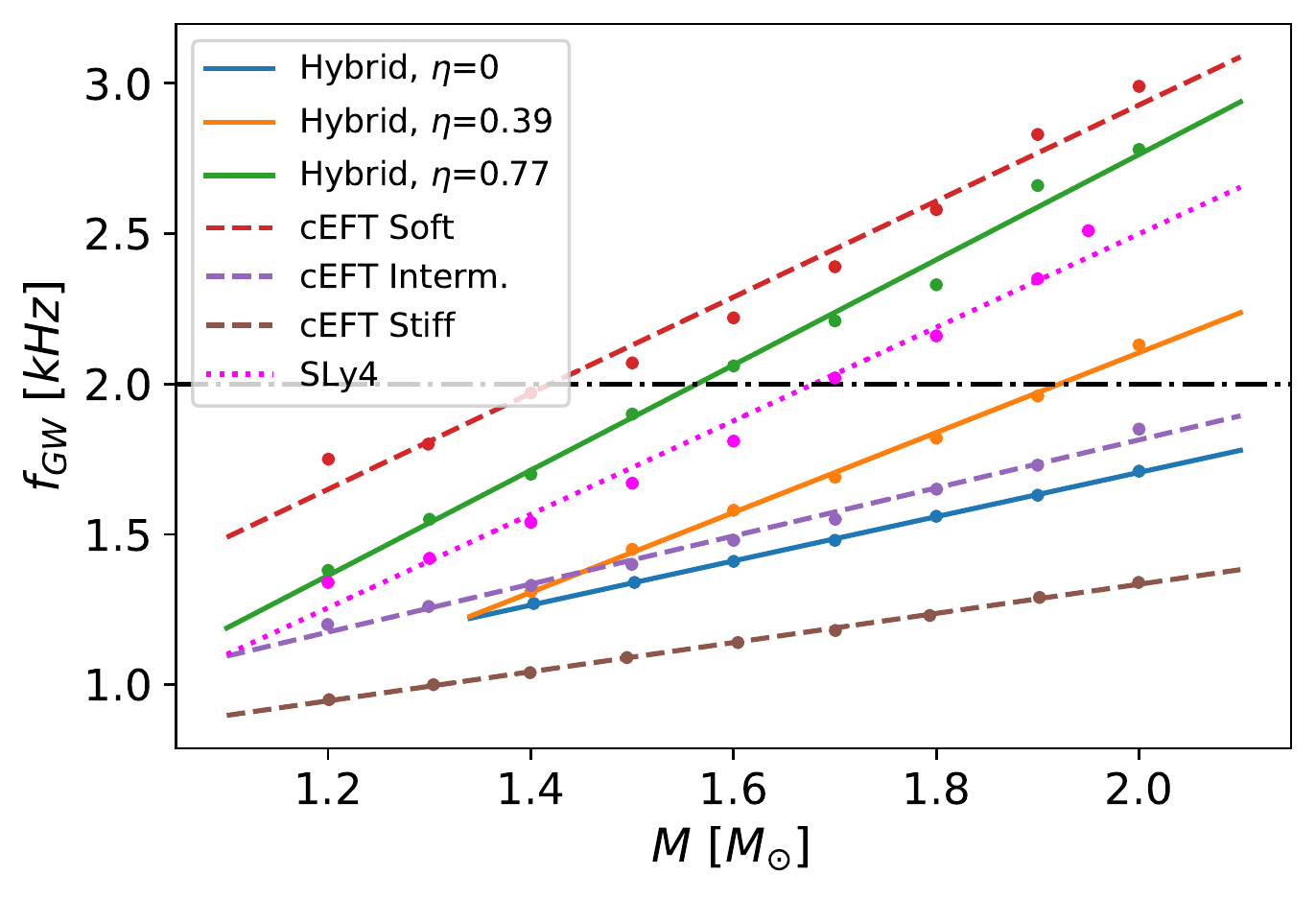}  
   \caption{Crust-breaking GW frequency of NSs in a binary system with a $1.4M_{\odot}$ companion for realistic EOSs. The shear modulus used in the calculations is a function of the EOS (see \citet{2008A&A...491..489Z}). A linear relationship between the critical frequency and the NS mass is evident for the mass range of astrophysical interest. It also holds for larger mass companions, the only difference being that it shifts to smaller values. The merger reference frequency (within the Newtonian context) of around 2 kHz concerns 1.4 $M_{\odot}$ stars with 10 km radii \citep{2020PhRvD.101j3025G}. We stress that for larger radii, such as the 12 km ones coming from most of the EOSs we have used, the merger frequency is lower.} For EOSs that are not too soft, many regions of the hadronic crust - not only the outermost ones - start fracturing before merger.
   \label{fig:Mfrealistic}
   \end{figure}

Given the generality of the above critical GW frequency for crust failure and NS mass, we expect a similar behavior for more realistic EOSs. Figure \ref{fig:Mfrealistic} shows that this is indeed the case for different NS models,  with a 1.4 $M_{\odot}$ companion in the binary. To obtain breaking frequencies associated with the failure of many parts of the crust and not only the shallower ones, we have searched for the lattice points that start yielding at higher densities, $\rho\gg\rho_{\rm{drip}}$ ($\rho>2\times 10^{13}$gcm$^{-3}$). Softer EOSs present a larger slope and higher breaking frequencies than stiffer ones. The reason is primarily due to the associated compactness of stars. Even cases with critical GW frequencies below 1kHz are possible, as is the case of stiff EOSs. For softer EOSs, there are cases where the crust does not even start failing before the merger, as previous analyses have already elucidated \citep{2020PhRvD.101j3025G}. This threshold depends on the NS mass and the companion's mass. For larger mass companions, the critical GW frequencies decrease for a given mass - the relationship between $f_{\rm {GW}}$ and $M$ still remains linear - which could be qualitatively understood as due to the increase of tidal forces on NSs and their subsequent more significant deformations. One can also see that the critical crust-breaking frequency directly relates to the stiffness of the EOS and traces well some of their $M(R)$ aspects, such as the points where the curves cross. Higher slopes of the crust-breaking frequency curves also relate to the stiffness of the EOSs. Therefore, observables associated with the crust yield might be able to independently constrain the stiffness of the EOS if the masses of the objects in the binary system are known with sufficient precision, as expected in many GW detections. 

A relevant aspect of Fig. \ref{fig:Mfrealistic} is that for a hybrid EOS NS, the crust-breaking GW frequencies depend on $\eta$. However, one also sees that hybrid NSs may behave similarly to purely hadronic NSs in terms of slopes. The main difference is a slope change associated with the phase transition in the case of several measurements. We elaborate on this point in Sec. \ref{sec:conclusions}.

\begin{figure} 
   \includegraphics[width=\columnwidth]{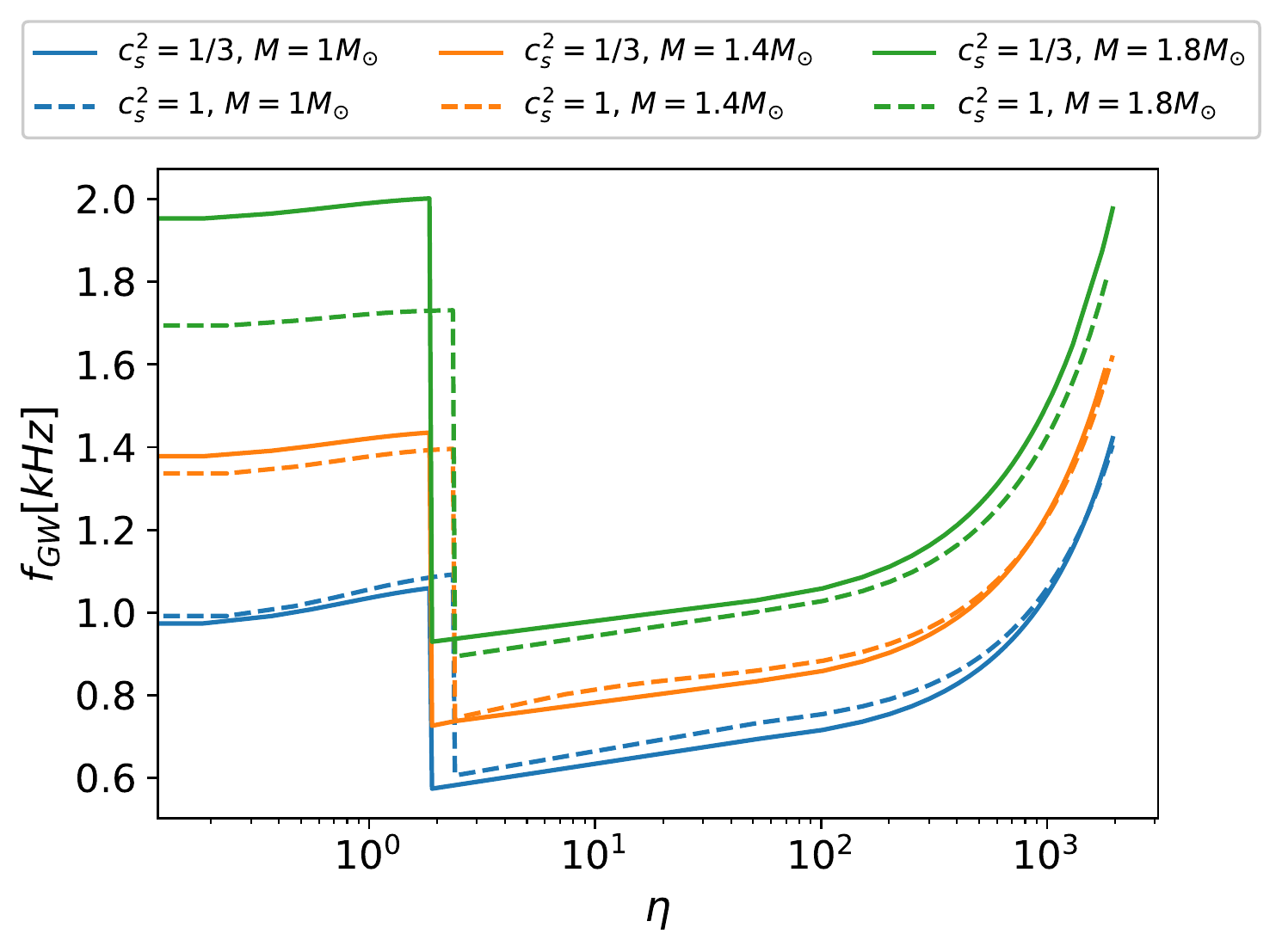}  
   \caption{Crust-breaking GW frequency for fixed NS masses and speeds of sound, with a binary companion of 1.4 $M_{\odot}$. The discontinuity refers to the case where the elastic hadronic phase directly touches the quark core. Aspects of the quark EOS are more relevant for heavier stars with small $\eta$. For other configurations, the critical crust-breaking frequency is only weakly dependent on the quark EOS.}\label{fig:frequency_jump_book}
   \end{figure}

Lets us consider the case of large density jumps by plotting the crust-breaking frequencies for fixed masses and speeds of sound. Figure \ref{fig:frequency_jump_book} shows the case of different masses for $\eta$ values as high as $10^3-10^4$ using the toy-model EOSs. The general trend is the increase of the crust-breaking frequency with $\eta$ due to the increase in compactness. One also notices a sharp discontinuity for the critical jump where the quark core starts touching the elastic hadronic phase. The reason for that is the change in the boundary conditions for tidal deformations (we elaborate on it in the sequel). Breaking frequencies for each NS mass have two representatives: small and large $\eta$ values. In the latter case, due to the higher compactness, the breaking frequencies are almost independent of the speed of sound for a given mass. The above means that one might constrain a set of parameters for phase transitions with crust-breaking frequency observations almost independently on the quark EOS. Notice, however, that the crust-breaking frequency depends more significantly on the NS mass, which could be inferred from GW measurements. To discriminate between small or large $\eta$, tidal deformations from the early inspiral could also help, given that for the toy-model EOSs investigated they  differ around $10\%-15\%$ for stars with a given mass in the most optimistic cases (when stars with very small and very large $\eta$s are compared). This difference suggests the level of accuracy for resolving the degeneracy, which could only be lifted with third-generation GW detectors (see, e.g.,
\citet{2023arXiv230405349P,2022PhRvL.128j1101P,2022PhRvD.105l3032W,2021PhRvL.127h1102S} and references therein).

\begin{figure} 
   \includegraphics[width=\columnwidth]{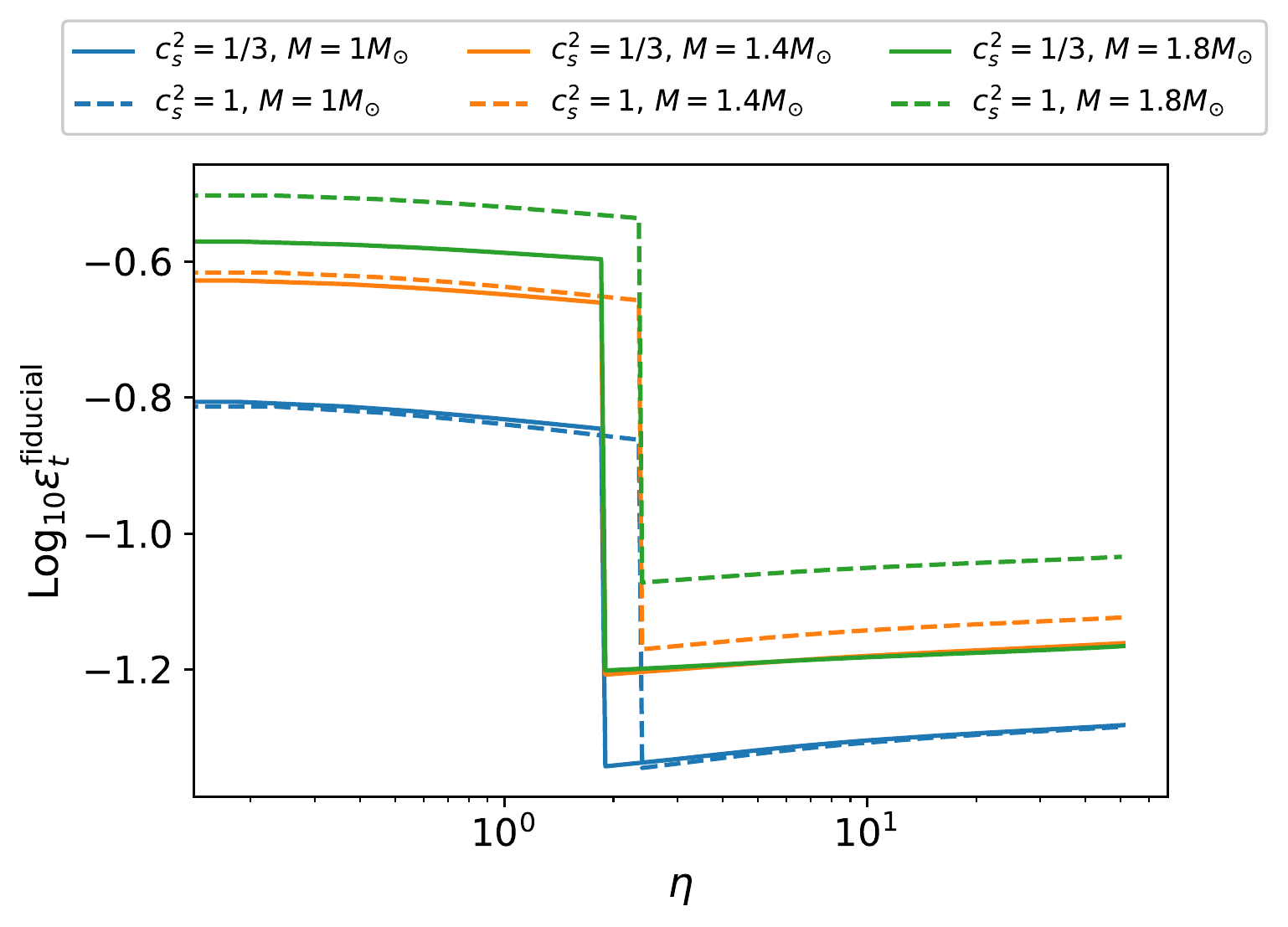}   
   \caption{Maximum ``total'' fiducial ellipticity of stars for different energy-density jumps in the presence of the tidal forces due to the binary companion of 1.4 $M_{\odot}$. The sudden change in the total ellipticity is uniquely due to a change in boundary conditions for perturbations when an elastic hadronic phase directly touches a liquid quark phase. }\label{fig:ellipticity_tidal_forces}
   \end{figure}

\begin{figure} 
   \includegraphics[width=\columnwidth]{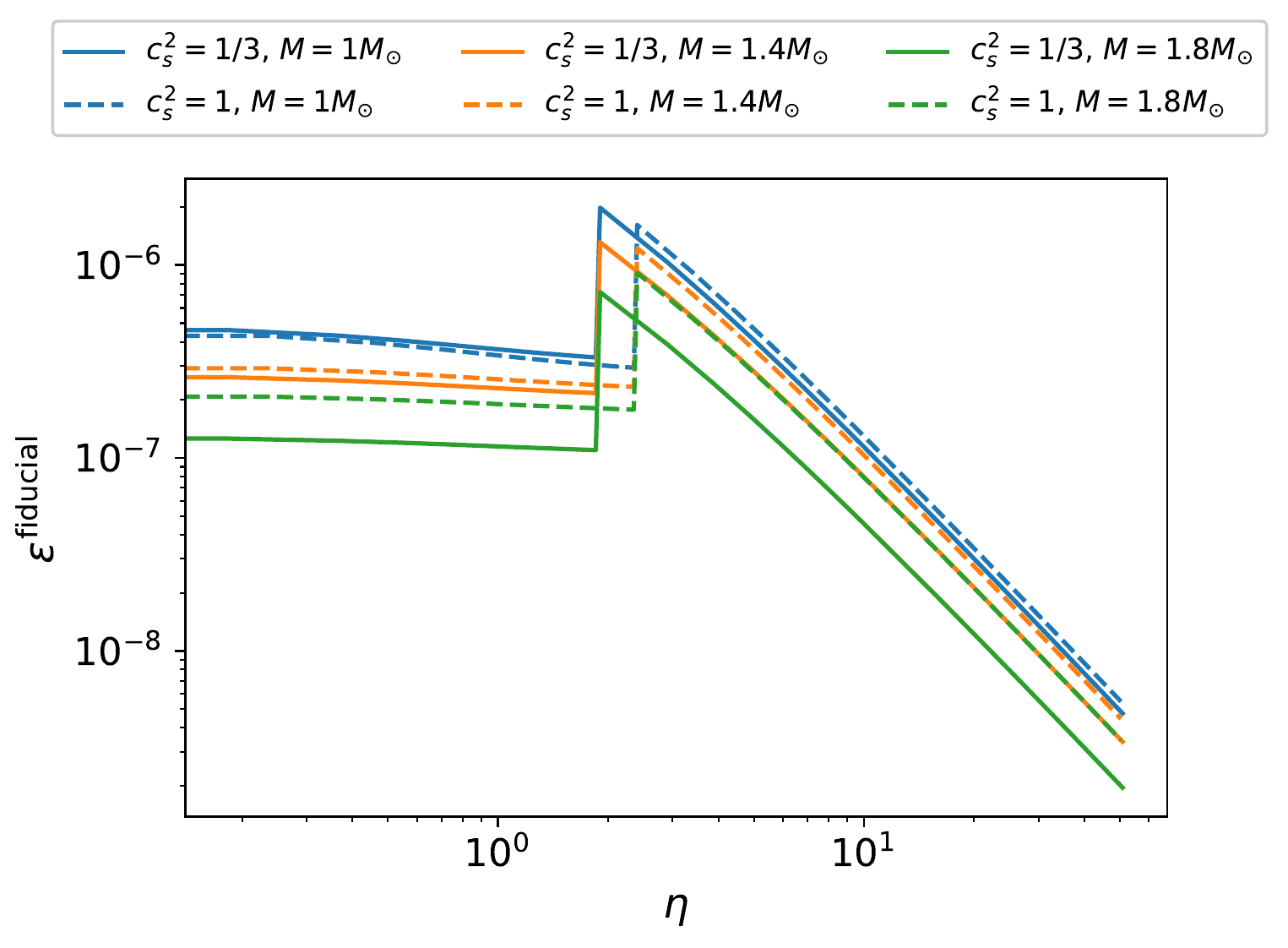}   
   \caption{Maximum ``net'' fiducial ellipticity of isolated NSs for different $\eta$'s when the force procedure of \citet{2021MNRAS.500.5570G} is taken into account. For an NS with a given mass, there is a range of maximum ellipticities with representatives for either small or very large energy-density jumps.}
   \label{fig:ellipticity}
   \end{figure}

The toy-model EOSs also allow us to have an idea about the properties of hybrid NS maximum ellipticities. In Fig. \ref{fig:ellipticity_tidal_forces}, we plot the maximum ellipticity of stars with elastic crusts in the presence of tidal forces. The maximum possible ``total'' ellipticity $(\varepsilon_t)$ is around $10^{-1}$, meaning a significant deformation of the NS. However, Fig. \ref{fig:ellipticity} shows how the ``net'' ellipticity, when the external force is subtracted off. For doing so, we have followed the procedure of \citet{2021MNRAS.500.5570G}, where the relaxed state of the star is a perfect fluid subjected to the same force as in the elastic case. That would be a way of calculating maximum ellipticities not violating any boundary condition and somewhat independently of the forces deforming NSs. They may be compared with the steadily improved upper limits on rotating NS ellipticities (non-axisymmetric deformations) obtained in the LIGO-Virgo-KAGRA searches for CWs. In the case of all-sky searches for \textit{a priori} unknown sources \citep{2022arXiv220100697T}, ranges for possible NS ellipticities as a function of the distance to the source are plotted, e.g. at the GW frequency of 1100 Hz, at a distance up to 1 kpc, possible ellipticities are smaller than $3\times 10^{-7}$ (see their left panel in Fig. 16). In the case of known NSs (pulsars) \citep{2022ApJ...935....1A}, several sources are already probed in the physically-interesting regime in which the spin-down observed in EM waves cannot be explained by the GW emission alone, i.e., the GW strain amplitude upper limit is below the spin-down limit: the upper limit for ellipticity of the Crab pulsar at the distance of about 2 kpc at the GW frequency of 59.2 Hz is about $8\times 10^{-6}$; the current lowest upper limit for ellipticity of $5.3\times 10^{-9}$ belongs to a millisecond pulsar J0711-6830 at a distance of 0.11 pc and GW frequency of 364.2 Hz (see their Fig. 4).  

Similarly to the case of the critical breaking frequencies in a binary NS system, the ellipticity would change abruptly (by around an order of magnitude) when the elastic crust directly touches the (liquid) quark phase. Larger values to the net ellipticity rise in this case. The reason is the sudden decrease of the total ellipticity in the elastic case, enhanced by the $\eta$ (which all decreases deformations), and the continuous behavior of the ellipticity only associated with the perfect-fluid reference star, which leads to a more considerable absolute difference. However, it quickly decreases when $\eta$ is large enough. It is also weakly dependent on the quark EOS.  

A discontinuous maximum ellipticity when a liquid quark core directly touches an elastic hadronic phase/elastic crust (which happens if the EOS presents an energy-density jump larger than a critical one, $\eta_{\rm{crit}}$, defined such that the density at the bottom of the hadronic phase is $\rho=1.5\times 10^{14}$ g cm$^{-3}$--the maximum density at the base of the elastic crust) deserves further analysis. For double-checking the consistency of our numerical calculations, we have made some extra tests. For doing so, we have used our toy-model EOS with $c_s^2=1$ for $M=1.4M_{\odot}$ hybrid stars. To assess the maximum ellipticity discontinuity, we have considered the cases where the energy density jump is slightly larger and smaller than $\eta_{\rm{crit}} \approx 2.352$. Table \ref{table_check} shows the results for relative changes of some observables and their best fits as a function of $\kappa_{\rho}$ when one assumes that $\check{\mu}=\kappa_{\rho}\rho$ and $\eta$s slightly larger and smaller than $\eta_{\rm{crit}}$. One can clearly see that relative changes go to zero linearly with $\kappa_{\rho}$ (the shear modulus). The fits are needed due to the high numerical fluctuations when $\check{\mu}\rightarrow 0$. The discontinuous nature of relative tidal deformations and maximum ellipticities when the liquid core touches either an elastic or a liquid hadronic phase stems from the boundary condition for $H_0'$. More explicitly, its jump at an interface is \citep{2020arXiv201106361P}
\begin{equation}
    [H_0']^+_-= 8\pi e^{\lambda(R_{i})}\frac{W_{\rm{le}}}{R_{i}}[\rho]^+_- -16\pi\check{\mu}^+ V^+\nu'\label{jumpofH00},
\end{equation}
where $e^{\lambda}$(-$e^{\nu}$) is the $rr$($tt$) component of the background metric, $R_{i}$ is the radius of a sharp interface splitting two phases, $W_{\rm{le}}$ and $V$ are related to the radial and angular volume element displacements at such interface (here the intersection of a liquid and an elastic (le) phase), and $``+"$($``-"$)  is to be understood as the elastic(liquid) phase immediately above(below) $R_i$, whose normal points outwards in the radial direction. Since the shear modulus of the hadronic crust is small (around $1\%$ of the pressure), it is usually the case that the second term of the above equation can be ignored when compared to the first one. When we have a sharp interface splitting two liquid phases, it follows that \citep{2020arXiv201106361P}
\begin{equation}
    W_{\rm{ll}}= \left[\frac{H_0 r^3}{2(m+4\pi r^3 p)}e^{-\lambda} \right]_{r=R_{i}}\label{W_ll},
\end{equation}
whereas in the case of a liquid-elastic interface, $W_{\rm{le}}$ is determined only through the fulfillment of the boundary conditions. In general, $W_{\rm{le}}$ and $W_{\rm{ll}}$ are different, and this difference shows itself in the relative tidal deformations as in Table \ref{table_check}. The fact that relative changes go to zero when the shear modulus vanishes means that $W_{\rm{le}}\rightarrow W_{\rm{ll}}$ only in this limit, as expected. We also note that when NS maximum ellipticities are concerned, the relativistic term $4\pi R_{i}^3 p(R_i)$ in the denominator of Eq. \eqref{W_ll} should not be ignored, as is usually the case for tidal deformation calculations \citep{2010PhRvD..82b4016P},  for the perfect-fluid reference case. Indeed, the continuity of the radial traction makes $W_{\rm{le}}$ reduce to Eq. \eqref{W_ll} in the perfect-fluid limit ($\check{\mu}\rightarrow 0$). Moreover,  
a perfect-fluid hybrid star in the presence of perturbations is our reference in the force-subtraction scheme for the maximum ellipticity we have used.

\begin{center}
\begin{table}
[htbp] 
\begin{ruledtabular}
{\begin{tabular}{@{}c|ccccc@{}} 
$\kappa_{\rho}$& $\left(\frac{\Lambda_{\rm{perf}}-\Lambda_{\rm{elas}}}{\Lambda_{\rm{perf}}}\right)_{\rm{le}}$ & $\left(\frac{\Lambda_{\rm{perf}}-\Lambda_{\rm{elas}}}{\Lambda_{\rm{perf}}}\right)_{\rm{ll}}$& $\varepsilon_{\rm{le}}^{\rm{fiducial}}$ & $\varepsilon_{\rm{ll}}^{\rm{fiducial}}$ \\ 
(cm$^2$s$^{-2})$& $(10^{-6}\%)$ & $(10^{-7}\%)$ &($10^{-10}$) & $(10^{-10})$  \\ \hline
$10^{16}$& $1818$ & $1061$ & $12267$  & $2334$ \\ $5.5\times 10^{15}$& $997.8$ & $583.7$ & $6751$  &  $1284$\\
$5.4\times 10^{14}$& $98.13$ & $57.37$ & $663.3$ & $126.2$\\ 
$10^{14}$& $18.18$ & $10.66$ & $122.87$ & $23.44$ \\ 
$10^{13}$& $1.823$ & $1.104$ & $12.322$ & $2.429$ \\  \hline
 Best Fit & $1.817\left(\frac{\kappa_{\rho}}{10^{13}}\right)$ & $1.061 \left(\frac{\kappa_{\rho}}{10^{13}}\right)$ & $12.269\left(\frac{\kappa_{\rho}}{10^{13}}\right)$ & $2.334\left(\frac{\kappa_{\rho}}{10^{13}}\right)$\\
\end{tabular}}
\end{ruledtabular}
\caption{Relative changes of tidal deformations (see text below Eq. \eqref{EOS_hadron} for details) and maximum fiducial ellipticities when the liquid quark core directly touches a liquid hadronic phase (ll) and when the liquid quark core directly touches an elastic hadronic phase (le) for stars with $M=1.4 M_{\odot}$ using our toy-model EOS with $c_s^2=1$. Differences in the relative changes are due to the effect of boundary conditions for $H_0'$ (see Eq.\eqref{jumpofH00}), which affect quantities such as the radial and tangential tractions and volume element displacements. As a result, the scalar strain in the elastic hadronic phase changes, and a discontinuity appears in the maximum ellipticity as a function of the energy-density jump. One can clearly see that the decrease of the shear modulus (through $\kappa_{\rho}$ by means of $\check{\mu}=\kappa_{\rho}\rho$) leads to a linear decrease in relative tidal deformations and maximum ellipticities such that they go to zero when $\kappa_{\rho} \rightarrow 0$, as expected.}
\label{table_check}
\end{table}
\end{center}

However, there are more subtleties when maximum ellipticities (and also crust-breaking frequencies) are taken into account when $\eta$ is around $\eta_{\rm{crit}}$ because one relies on the scalar strain ($\Theta$, see Eq. \eqref{scalar_strain}) for obtaining them, which depends on volume element displacements, perturbation spacetime functions and tractions in the elastic phase. The different boundary conditions for the case of liquid-liquid ($\eta$ is slightly smaller than $\eta_{\rm{crit}}$) and liquid-elastic ($\eta$ is slightly larger than $\eta_{\rm{crit}}$) interfaces influence nontrivially these perturbation quantities throughout the crust. For our toy-model EOSs, the scalar strain is maximum at the base of the elastic crust but it has different strengths for different boundary conditions. In the toy-model EOS with $c_s^2=1$
for a $1.4 M_{\odot}$ star and an arbitrary perturbation amplitude, the scalar strain is approximately three times larger throughout the crust when the liquid core touches an elastic crust than when it touches a liquid crust. This difference arises for the same shear modulus throughout the crust because we are just considering $\eta$s around $\eta_{\rm{crit}}\approx 2.352$. As a result, both the total ellipticity due to tidal forces and the maximum ellipticity of isolated NSs exhibit discontinuous behavior. The total ellipticity is smaller when $\eta$ is slightly larger than $\eta_{\rm{crit}}$  (see Fig. \ref{fig:ellipticity_tidal_forces}) because, due to the three-times larger crustal scalar strain in this case, three-times smaller perturbation amplitudes are needed to reach the breaking strain. On the other hand, the maximum ellipticity increases when $\eta$ is slightly larger than $\eta_{\rm{crit}}$ because, in the force-subtraction procedure, the contribution from the reference liquid hybrid star is smaller due to the need for a smaller maximum perturbation amplitude.     

We should also estimate the errors for the crust-breaking GW frequencies coming from our Newtonian approximations. We start with Kepler's third law. Eq. (228) of \citet{2014LRR....17....2B} tells us that the relative uncertainty for the angular frequency due to the 1PN (relativistic) corrections is
\begin{equation}
    \frac{\Delta \Omega}{\Omega}\equiv \frac{|\Omega-(M_t/r^3)^{\frac{1}{2}}|}{(M_t/r^3)^{\frac{1}{2}}}\simeq \frac{1}{2}(3-\nu)(\pi M_t f_{\rm {GW}})^{\frac{2}{3}}\label{omega_error},
\end{equation}
where  $M_t\equiv M_1+M_2$ is the total mass of the binary and $\nu\equiv M_1M_2/M_t^2$. For an equal-mass binary with $1.4M_{\odot}$, a crust-breaking GW frequency ($f_{\rm {GW}}$) around (1-2)kHz (see Fig. \ref{fig:Mfrealistic}) would imply that $\Delta f_{\rm {GW}}/f_{\rm {GW}}\simeq \Delta \Omega/\Omega \simeq 15\%-30\%$. 
For comparison, in the optimistic case where $f_{\rm {GW}}=500$ Hz (slightly below the smallest crust-breaking GW frequency of Fig. \ref{fig:frequency_jump_book}, approximately 600 Hz), one would have $\Delta f_{\rm {GW}}/f_{\rm {GW}} \simeq 10\%$. Kepler's third law also needs to be corrected due to tidal deformations, as shown in Eq. (A6) of \citet{2014PhRvD..89d3011Y}. However, for $\Lambda_{1.4M_{\odot}}\simeq 500$ and crust-breaking GW frequencies within (1-2)kHz, the relative changes of $f_{GW}$ are very small ($\Delta f_{\rm {GW}}/f_{\rm {GW}}\simeq 10^{-2}\% - 10^{-1}\%$). Our crust-breaking GW frequency calculations use the Newtonian quadrupolar tidal field (${\cal E}_{ab}$), so it is necessary to estimate errors due to relativistic corrections. When 1PN corrections to the tidal field are taken into account, from  Eqs. (9.3) and (9.13) of \citet{2018PhRvD..97l4048P}, we have that
\begin{equation}
    \frac{\Delta {\cal E}_{ab}}{{\cal E}_{ab}}\equiv \frac{|{\cal E}_{ab}-{\cal E}_{ab}(\rm{0PN})|}{{\cal E}_{ab}(\rm{0PN})}\sim \left(1-\frac{M_1}{2M_t} \right)(\pi M_t f_{\rm {GW}})^{\frac{2}{3}},
\end{equation}
where $M_1$ is the body's mass creating the tidal field.
For GW frequencies in the range $(1-2)$kHz and equal-mass binaries with $1.4M_{\odot}$, we have $\Delta {\cal E}_{ab}/{\cal E}_{ab}\simeq 10\%-20\%$. Since ${\cal E}_{ab}$ is proportional to $f_{\rm {GW}}^2$ \citep{2020PhRvD.101j3025G}, it follows that $\Delta f_{\rm {GW}}/f_{\rm {GW}}\sim (1/2)\Delta {\cal E}_{ab}/{\cal E}_{ab} \simeq 5\%-10\%$.
From the above, uncertainties are not small in general, meaning that for precise calculations relativistic corrections to Kepler's third law and the tidal field should not be ignored. However, even closer to merger, they are still much less than the GW frequency uncertainties associated with different EOS possibilities.

\section{Discussion and conclusions}
\label{sec:conclusions}

GW waveforms are not expected to change significantly due to the crust cracking because the breaking frequencies are large, meaning that the NSs only have a few cycles before merger. One can roughly estimate the change based on the analysis of \citet{1994MNRAS.270..611L,2020PhRvL.125t1102P} when adapted to the static case. For 1.3 $M_{\odot}$--the reference value of \citet{2020PhRvL.125t1102P}--, the minimum crust-breaking frequency is around 750 Hz (see Fig. \ref{fig:Mfrealistic}), and the elastic energy is around $10^{47}$ erg. Inserting these parameters into Eq. 9 in \citet{2020PhRvL.125t1102P}, one has that the largest GW phase shift change is $\delta\phi \propto (f_{\rm{GW}}/70\,\mathrm{Hz})^{-7/3}\sim 4\times 10^{-4}$. Dynamical tides, in addition to changing static tides by a few percent \citep{2020PhRvD.101h3001A}, may have a more pronounced effect on the process of crustal breaking and may lead to observable effects \citep{2020PhRvL.125t1102P,2021MNRAS.504.1273P}. We leave dynamical tide studies of hybrid NSs for future work.

One can estimate the amount of elastic energy released during the crustal failure. Roughly, it is $10^{-2}\int\check{\mu} dV$ \citep{2012ApJ...749L..36P,2012PhRvL.108a1102T}. For $\eta\approx 10^3$, the typical energies are of the order of $10^{39}-10^{40}$ erg. For $\eta \lesssim 1$, the elastic energy is smaller than $10^{46}-10^{47}$erg. One would expect the release of elastic energy during the plastic regime through a cascade effect \citep{2020PhRvL.125t1102P}. How elastic energy converts into other forms of energy is however not fully understood. EM radiation could appear if the crust breaking is dynamical, and if it excites seismic waves in the crust. These waves could transform into shock waves due to the strong density gradient near the crust surface, at the bottom of the star's magnetosphere \citep{1990ApJ...363..612B}.
An excited magnetosphere would then produce a short EM radiation burst--a (weak) precursor of a powerful short gamma ray burst associated with an NSNS merger (see, e.g., \citet{2020ApJ...902L..42W,2020PhRvD.102j3014C,2021Galax...9..104W} and references therein). Another part of the elastic energy may go into heat; however, it may not be enough to melt the crust of the star \citep{2012ApJ...749L..36P}.

Crust-breaking frequencies scaling linearly with masses could have significant consequences for independent measurements for NS observables, in addition to the ones inferred from GWs. Depending on the orbital/GW frequency associated with the EM precursor, one could independently constrain the EOS if the NS mass is known, which is expected with some error uncertainty from the GW waveforms. In addition, if at least two observations are made around similar NS masses (for similar companion masses), and the results are very different (larger than uncertainties), this would point to the critical mass associated with a phase transition. If several measurements are made for a similar companion mass, a change in slope of the crust-breaking frequency as a function of the mass of the NS would also suggest a phase transition inside the star. Indeed, for a one-phase EOS, irrespective of its properties, one would expect no change of slope. However, it still remains challenging to precisely disentangle a hybrid NS EOS from a one-phase one based on crust-breaking frequencies in general cases. That is due to their dependence on density jumps, stiffnesses for a given class of phase transitions (small or substantial density jumps), and even companion masses.

Interestingly, Fig. \ref{fig:frequency_jump_book} suggests that a given GW frequency for the precursor would lead to either very small or large energy density jumps for a star with a given mass. One could probe the correct case with tidal deformation measurements because they depend on density jumps (see, e.g., \citet{2020arXiv200310781P} and references therein). However, relative tidal deformation differences for both cases would differ around 10$\%$-15$\%$ in the most optimistic cases, rendering such measurements possible only for rare high signal-to-noise events or third-generation detectors (see \citet{2022PhRvD.105l3015P} and references therein). Even if no precursor observations are made, one could still infer some independent aspects of the NS interior, suggesting a soft EOS.

Concerning the maximum quadrupolar deformation of hybrid NSs, their dependence on $\eta$ may also shed some light on the phase transitions of dense matter. Figure \ref{fig:ellipticity} suggests that isolated NSs with a given maximum ellipticity may have very small or intermediate energy-density jumps. In addition, a liquid quark core directly touching an elastic hadronic phase--related to a phase transition with a large quark-hadron energy-density jump--could increase the maximum ellipticity of a star by around an order of magnitude when compared to the case of a liquid quark core touching a liquid hadronic phase. This is potentially relevant for continuous GW detections (since it would increase the odds of detecting them) and hence it should be further investigated. Upper limits set by the LIGO-Virgo-KAGRA collaboration for several known pulsars \citep{2022ApJ...935....1A} may already constrain the parameter space of energy-density jumps, especially the ones with small and very large energy-density jumps, which have smaller maximum ellipticities. However, one should bear in mind many caveats for direct comparisons. First, evolutionary history should play an essential role in the maximum ellipticity of a star, and it is not taken into account in theoretical investigations at all so far. Second, isolated NSs usually have unknown masses, and the maximum ellipticities may change an order of magnitude depending on them. Third, when the sensitivity increases, the upper limits obtained by the non-detection of CWs will be more constraining. Fourth, upper limits to ellipticities of pulsars are not a direct measure of their maximum possible deformations. They are, as of now, a guide for theoretical models taking into account various ingredients and assumptions. For example, if they predict values that are smaller than measured ones, then they suggest further relevant elements are lacking. However, as our toy models indicate, phase transition aspects could allow larger maximum ellipticities for intermediate energy-density jumps when compared to small ones, meaning that there is still some space of parameters to be investigated and only further measurements and statistical analysis will be able to constrain the internal structure of stars in general. A more strict constraint could also happen, in particular, if a serendipitous measurement of a pulsar with a very large ellipticity were measured because it could exclude many models.

Lastly, we stress that have mostly focused on order-of-magnitude estimates and general trends for crust-breaking frequencies and maximum ellipticities. Precise numbers to be directly compared with GW data would require much more accurate analysis and models. For instance, post-Newtonian corrections to Kepler's third law and the tidal field should not be ignored for crust-breaking frequencies, and realistic hadronic EOSs calculations are needed for maximum quadrupolar deformations (ellipticities). The role played by intermediate energy-density jumps (when a quark core directly touches an elastic hadronic phase) should also be analyzed with more detail because it can lead to larger deformations and smaller crust-breaking frequencies, which are clearly relevant for continuous GW detections and earlier-in-the-coalescence EM phenomena. Finally, the impact of a mixed phase on crust-breaking frequencies and maximum ellipticities also needs special attention, given that it is a possibility in hybrid stars and it could allow large quadrupolar deformations.

Summing up, we found that internal aspects of NSs such as phase transitions can leave an imprint on their maximum ellipticities and crust-breaking frequencies in binary systems. Crust-breaking frequencies scale linearly with the NS mass for a given mass of the companion, and the slope of this relation depends on the stiffness of the EOS, on which phase transition properties also depend. Maximum ellipticities for hybrid NSs with small $\eta$ behave similarly to one-phase stars but when the $\eta$ values are intermediate, hybrid NSs with elastic crusts may deform in a more pronounced way. A given maximum ellipticity or crust-breaking frequency would have two possibilities within hybrid NSs: the associated density jumps are either small or intermediate. That may be disentangled with tidal deformation measurements but the precision required is estimated to be around $10\%-15\%$ in the most optimistic cases, which will be possible with third-generation GW detectors. Finally, more precise models and analysis are needed if one wishes to directly compare crust-breaking outcomes of hybrid stars with GW and EM observations.

\section{Acknowledgments}
We thank Andrew Melatos for useful comments and the anonymous referee for very useful suggestions which have improved our work. The Authors gratefully acknowledge the financial support of the National Science Center Poland grants no. 2016/22/E/ST9/00037, 2018/29/B/ST9/02013 and 2021/43/B/ST9/01714, and the Italian Istituto Nazionale di Fisica Nucleare (INFN), the French Centre National de la Recherche Scientifique (CNRS) and the Netherlands Organization for Scientific Research (NWO), for the construction and operation of the Virgo detector and the creation and support of the EGO consortium. 

\bibliographystyle{aasjournal}
\bibliography{bibliography_mountains}

\end{document}